# On the absolute value of the neutrino mass

Dragan Slavkov Hajdukovic[1]
PH Division CERN
CH-1211 Geneva 23;
dragan.hajdukovic@cern.ch
[1]On leave from Cetinje; Montenegro

*Abstract*
The neutrino oscillations probabilities depend on mass squared differences; in the case of 3-neutrino mixing, there are two independent differences, which have been measured experimentally. In order to calculate the absolute masses of neutrinos, we have conjectured a third relation, in the form of a sum of squared masses. The calculated masses look plausible and are in good agreement with the upper bounds coming from astrophysics.

The recent investigations of neutrinos from the sun and of neutrinos created in the atmosphere by cosmic rays, have given strong evidence for neutrino oscillations, i.e. phenomenon when neutrinos change from one flavour to another (See Ref. 1 and 2 for a recent Review). Neutrino flavour change implies that neutrinos have masses. To determine these masses remains one of the most challenging tasks of contemporary physics, bearing fundamental implications to particle physics, astrophysics and cosmology. In the present paper we have calculated the absolute mass of neutrinos by using the existing experimental data and a new theoretical assumption.

Let's consider the case when there are three neutrino flavour eigenstates $(\nu_e, \nu_\mu, \nu_\tau)$ and three neutrino mass eigenstates $(\nu_1, \nu_2, \nu_3)$ with corresponding masses $(m_1, m_2, m_3)$. The neutrino oscillations probabilities depend on mass squared differences, $\Delta m_{ij}^2 \equiv (m_i^2 - m_j^2)$, which can be determined experimentally. In the case of 3-neutrino mixing there are only two independent neutrino mass squared differences, say $\Delta m_{21}^2$ and $\Delta m_{32}^2$. The $\Delta m_{21}^2$ (needed to explain solar data) and $\Delta m_{32}^2$ (needed to explain the atmospheric data) are often referred as the "solar" and "atmospheric" neutrino squared differences and denoted as $\Delta m_{21}^2 \equiv \Delta m_{Sun}^2$ and $\Delta m_{32}^2 \equiv \Delta m_{atm}^2$.

From the recent experiments, we know [1] that:

$$\Delta m_{Sun}^2 \equiv \Delta m_{21}^2 \equiv m_2^2 - m_1^2 \approx 7.59 \times 10^{-5} eV^2 \qquad (1)$$

$$\left|\Delta m_{atm}^2\right| \equiv \left|\Delta m_{23}^2\right| = \left|m_3^2 - m_2^2\right| \approx 2.43 \times 10^{-3} eV^2 \qquad (2)$$

Hence, neutrino spectrum contains a doublet of two mass eigenstates ($m_1$ and $m_2$) separated by the splitting $\Delta m_{Sun}^2$, and a third eigenstate $m_3$, separated from the first two by a much larger splitting $\Delta m_{atm}^2$ ($\Delta m_{atm}^2 \approx 32 \Delta m_{Sun}^2$).

As the equation (2) indicates, the existing data do not allow one to determine the sign of $\Delta m_{atm}^2$. The two possible signs of $\Delta m_{atm}^2$ correspond to two types of neutrino mass spectrum: spectrum with normal hierarchy ($m_1 < m_2 < m_3$) and spectrum with inverted hierarchy ($m_3 < m_1 < m_2$).

The two equations (1) and (2) are obviously not sufficient to determine three masses $m_1, m_2, m_3$. A third relation between masses is needed. For instance, a recent proposal [3] is to use, as a third equation, geometric mean neutrino mass relation $m_2 = \sqrt{m_1 m_3}$ for normal spectrum and $m_1 = \sqrt{m_2 m_3}$ for the inverted spectrum. This proposal is inspired by the geometric mean mass relation used previously for quarks. A weak point of the proposal lies in the well known fact [1, 2] that leptonic mixing (characterized with large mixing angles) is very different from its quark counterpart, where all the mixing angles are small. Our second objection is that geometric mean mass relation has a form which is quite different from Equations (1) and (2) while it is desirable to have a third equation with similar form. The equations (1) and (2) are about mass squared differences; hence the most natural third equation is in the form of a mass squared sum. So, as a third equation we propose

$$m_1^2 + m_2^2 = am_3^2; \quad 0 < a < 2 \tag{3}$$

for normal spectrum and

$$m_1^{'2} + m_3^{'2} = bm_2^{'2}; \quad 0 < b < 2 \tag{4}$$

for inverted spectrum. We have denoted masses of inverted spectrum by a prime. From the mathematical point of view, for sure, there is a constant $a$ (or $b$) for which the above relation is exact. At the end of this Letter and in the Appendix we would try to give a physical meaning to the above mass squared sums.

Equations (1), (2), (3) and (4) lead to

$$m_3^2 = \frac{2}{2-a}\Delta m_{32}^2 + \frac{1}{2-a}\Delta m_{21}^2 \tag{5}$$

$$m_2^2 = \frac{a}{2-a}\Delta m_{32}^2 + \frac{1}{2-a}\Delta m_{21}^2 \tag{6}$$

$$m_1^2 = \frac{a}{2-a}\Delta m_{32}^2 + \frac{a-1}{2-a}\Delta m_{21}^2 \tag{7}$$

for normal hierarchy, and

$$m_2^{'2} = \frac{1}{2-b}\left|\Delta m_{32}^2\right| + \frac{1}{2-b}\Delta m_{21}^2 \tag{8}$$

$$m_1^{'2} = \frac{1}{2-b}\left|\Delta m_{32}^2\right| + \frac{b-1}{2-b}\Delta m_{21}^2 \tag{9}$$

$$m_3^{'2} = \frac{b-1}{2-b}\left|\Delta m_{32}^2\right| + \frac{1}{2-b}\Delta m_{21}^2 \tag{10}$$

in the case of inverted hierarchy.

In order to calculate neutrino masses we must determine $a$ and $b$. The most elegant possibility (but we do not know if nature has the same aesthetic taste as we have) is $a = 1$ in the normal spectrum and $b = 1$ in the inverted spectrum; what reduces the equation (3) and (4) to the form of Pythagoras theorem. In addition to the beauty, the choice $a = 1$ and $b = 1$ assures that normal and inverted spectrum differ only in the third eigenstate ($m_3 > m_3^{'}$), while the other two eigenstates have the same value in both spectrums (i.e. $m_1 = m_1^{'}$ and $m_2 = m_2^{'}$). We consider it as a hint in favour of this choice.

The choice $a = 1$ and $b = 1$, together with the experimental results in equations (1) and (2), leads to the following absolute neutrino masses:

$$m_3 = 0.070 eV$$
$$m_2 = m_2^{'} = 0.050 eV$$
$$m_1 = m_1^{'} = 0.049 eV \qquad (11)$$
$$m_3^{'} = 0.0087 eV$$

In principle the masses obtained above must be checked against known experimental constraints, which unfortunately still have no a satisfactory accuracy. The best we can do is to compare with the upper limit on the sum of neutrino masses, deduced from astrophysical observations. According to (11) the sum of neutrino masses is about $0.17 eV$ for normal and $0.10 eV$ for inverted spectrum what is in agreement with the tightest current bounds, like a recent upper bound of $0.28 eV$ [4] determined from a photometric galaxy redshirt survey.

The physical ground for relations (1) and (2) is the fact that the neutrino oscillations probabilities depend on mass squared differences. What could be physical ground for mass squared sums (3) and (4)? We suggest that mass squared sums are related to a hypothetical source of neutrinos, characterized with the following emission probabilities

$$w_i = A m_i^2 ; \quad \sum_{i=1}^{3} w_i = 1 \qquad (12)$$

where $A$ is a constant depending on the physical characteristics of the source. Then, the equations (3) for the normal spectrum follows from (12) with $a$ being

$$a = \frac{w_1 + w_2}{w_3} \qquad (13)$$

and

$$w_1 + w_2 = \frac{a}{a+1} ; \quad w_3 = \frac{1}{a+1} \qquad (14)$$

It is easy to obtain the analogous results for inverted spectrum.

Now, according to (13), the choice $a = 1$ may be interpreted as equal emission probability for a "doublet" ($w_1 + w_2$) and a "singlet" ($w_3$).

A source of neutrinos with emission probabilities (12) is not known in nature, but it is striking that assumption of such a source leads to a plausible absolute mass of neutrino. In Appendix, we would argue that such a source is possible in the framework of contemporary physics.

**Appendix**

Let us remember the Schwinger mechanism [5] in Quantum Electrodynamics: a strong electric field $E$ greater than a critical value $E_{cr}$, can create electron-positron pairs (or other charged particle-antiparticle pairs) from the quantum vacuum. Hence, an external field can separate virtual particle and antiparticle, i.e. transform a virtual pair into a real one. In the limit $E >> E_{cr}$, the exponential factor in Schwinger relation [5] becomes a constant, and the particle-antiparticle creation rate per unite volume and time is proportional to the squared mass of considered particles.

$$\frac{dN_{m\bar{m}}}{dtdV} = B m^2 \qquad (15)$$

Consequently, according to (12) and (15), the masses of the charged leptons (electron, muon and tau) must satisfy a relation of the form (3). However, neutrinos are not charged particles and the original Schwinger mechanism does not work for them.

One interesting possibility to get relation (15) for neutrinos is to assume the existence of the gravitational repulsion between matter and antimatter [6, 7, 8]. If so, the relations (15), and consequently the sum (3), are valid for neutrinos through a gravitational version [6, 7] of the Schwinger mechanism.

As it was argued in [6] a sufficiently strong gravitational field needed for validity of relation (15) may exist only deep inside the horizon of a black hole, from where the created antineutrinos, are violently ejected. Hence, a black hole might behave as a point like source of antineutrinos. In principle, the study of antineutrino radiation of supermassive black holes in the centre of Milky Way and Andromeda Galaxy may confirm this phenomenon and lead to determination of probabilities (14) and the constant $a$. The study of solar neutrinos was crucial for relation (1): it is a striking speculation that the study of black hole neutrinos might be crucial for relation (3).